\documentclass[12pt]{article}

\renewcommand{\baselinestretch}{1.65}

\newcommand{\rA}{\rightarrow}

\newcommand{\xs}{\shortstack}

\begin{document}

\title{THE LEVI-CIVITA SPACETIME AS A LIMITING CASE OF THE $\gamma$
SPACETIME}
\author{L. Herrera$^1$\thanks{Postal address: Apartado 80793, Caracas 1080A,
Venezuela; e-mail: laherrera@telcel.net.ve}
,\ \
Filipe M. Paiva$^2$\thanks{e-mail: fmpaiva@symbcomp.uerj.br}
\\ \ \ and \ \
N. O. Santos$^3$\thanks{e-mail: nos@lacesm.ufsm.br} 
\\ \\ 
{\small $^1$Escuela de F\'{\i}sica, Facultad de Ciencias,}\\
{\small Universidad Central de Venezuela, Caracas, Venezuela, and}\\
{\small Centro de Astrof\'{\i}sica Te\'orica, Merida, Venezuela.}
\\ \\ 
{\small $^2$Departamento de F\'{\i}sica Te\'orica, Universidade do Estado do
Rio de Janeiro,}\\
{\small Rua S\~ao Francisco Xavier 524, 20550-013 Rio de Janeiro - RJ,
Brazil.}
\\ \\
{\small $^3$Laborat\'orio de Astrof\'{\i}sica e Radioastronomia,}\\
{\small Centro Regional Sul de Pesquisas Espaciais - INPE/MCT}\\
{\small Cidade Universit\'aria, 97105-900 Santa Maria RS, Brazil.}}
\date{}

\renewcommand{\baselinestretch}{1.1}
\maketitle

\begin{abstract}
It is shown that the Levi-Civita metric can be obtained from a family of the
Weyl metric, the $\gamma$ metric, by taking the limit when the length of its
Newtonian image source tends to infinity. In this process a relationship 
appears between two fundamental parameters of both metrics.
\end{abstract}
%{\sc pacs} numbers: 04.20.-q \ \ 04.20.Cv \ \ 04.20.Jb

\newpage

\section{Introduction} \label{SecInt} \setcounter{equation}{0}

One of the most interesting metrics of the family of
Weyl solutions \cite{Weyl} is the so called $\gamma$ metric
\cite{EspositoWitten1975,Virbhadra}. This metric, which is also known
as Zipoy-Voorhees metric \cite{Bach}, is continuously linked to the
Schwarzschild spacetime through one of its parameters and corresponds to a
solution of the Laplace equation in cylindrical coordinates. 
Its Newtonian image source \cite{Bonnor} is given by a finite
rod of matter. For a particular value of the mass density of the rod,
the metric becomes spherically symmetric (Schwarzschild metric).

In this article we show that extending the length of the rod to infinity
we obtain the Levi-Civita spacetime. At the same time a link is
established between the parameter $\gamma/2$, measuring the mass density
of the rod in the $\gamma$ metric, and the parameter $\sigma$, which is
thought to be related to the linear energy density of the source of the
Levi-Civita spacetime \cite{Bonnor}. Since $\sigma$ is the real source,
not the Newtonian image source and $\gamma/2$ measures the line mass density
of the Newtonian image source, not of the real source, our result
illustrates further the difficulties appearing in the interpretation of
the Levi-Civita metric as representing an infinite line mass of density
$\sigma$ \cite{Bonnor1}.

In the next section we describe the $\gamma$ metric. In
section~\ref{SecLimits} we show that it has a limit on the Levi-Civita
spacetime. In section~\ref{SecDiagram} some other limits are studied in order
to build a limiting diagram for the $\gamma$ metric. Finally 
section~\ref{SecConc} presents our conclusions.

\section{The $\gamma$ metric} \label{Secgamma} \setcounter{equation}{0}

In cylindrical coordinates, static axisymmetric solutions to Einstein's
equations are given by the Weyl metric \cite{Weyl}
\begin{equation} \label{gamma-metric-ds2}
ds^2=e^{2\lambda}dt^2-e^{-2\lambda}[e^{2\mu}(d\rho^2+dz^2)+
\rho^2d\phi^2],
\end{equation}
with
\begin{equation}
\lambda_{,\rho\rho}+\rho^{-1}\lambda_{,\rho}+\lambda_{,zz}=0,
\end{equation}
and
\begin{eqnarray}
\mu_{,\rho}&=&\rho(\lambda^2_{,\rho}-\lambda^2_{,z}),\\
\mu_{,z}&=&2\rho\lambda_{,\rho}\lambda_{,z},
\end{eqnarray}
where a comma denotes partial derivation. Observe the most amazing fact, as
Synge writes \cite{Weyl}, that (2.2) is just the Laplace equation for
$\lambda$ in Euclidean space.
The $\gamma$ metric is defined by \cite{EspositoWitten1975}
\begin{eqnarray}
e^{2\lambda}&=&\left[\frac{R_1+R_2-2m}{R_1+R_2+2m}\right]^\gamma,
\label{gamma-metric-lambda} \\
e^{2\mu}    &=&\left[\frac{(R_1+R_2+2m)(R_1+R_2-2m)}{4R_1R_2}\right]^
{\gamma^2}, \label{gamma-metric-mu}
\end{eqnarray}
where
\begin{equation} \label{gamma-metric-R1R2}
R_1^2=\rho^2+(z-m)^2, \qquad R_2^2=\rho^2+(z+m)^2.
\end{equation}
It is worth noticing that $\lambda$ as given by (2.5) corresponds to the
Newtonian potential of a line segment of mass density $\gamma/2$ and length
$2m$, symmetrically distributed along the $z$-axis. The particular case
$\gamma=1$ corresponds to the Schwarzschild metric. This is more easily seen
using Erez-Rosen coordinates \cite{Bach}, given by
\begin{equation} \label{cil-esf}
\rho^2=(r^2-2mr)\sin^2\theta, \qquad z=(r-m)\cos\theta,
\end{equation}
which yields the line element \cite{EspositoWitten1975}
\begin{equation} \label{gamma-esf}
ds^2=Fdt^2-F^{-1}[Gdr^2+Hd\theta^2+(r^2-2mr)\sin^2\theta d\phi^2],
\end{equation}
where
\begin{eqnarray}
F&=&\left(1-\frac{2m}{r}\right)^{\gamma},\\
G&=&\left(\frac{r^2-2mr}{r^2-2mr+m^2\sin^2\theta}
\right)^{\gamma^2-1},\\
H&=&\frac{(r^2-2mr)^{\gamma^2}}{(r^2-2mr+m^2\sin^2\theta)^{\gamma^2-1}}.
\end{eqnarray}
Now, it is easy to check that $\gamma=1$ corresponds to the Schwarzschild
metric.
The total mass of the source is $M=\gamma m$ \cite{EspositoWitten1975,Virbhadra}, and
its quadrupole moment $Q$ is given by
\begin{equation}
Q=\frac{\gamma}{3}M^3(1-\gamma^2).
\end{equation}
So that $\gamma>1$ ($\gamma<1$) corresponds to an oblate (prolate) spheroid.
We shall now show that {\it elongating} the Newtonian image source to
infinity we obtain the Levi-Civita spacetime. To achieve that, use will be
made of the Cartan scalars. In the next section these scalars are obtained
for the $\gamma$ metric, and are compared to the corresponding quantities of
the Levi-Civita metric in the limit $m\rightarrow\infty$.

\section{The Levi-Civita limit} \label{SecLimits} \setcounter{equation}{0}

Since the limit $m\rA\infty$ taken on the $\gamma$ metric in the form
(\ref{gamma-metric-ds2}) diverges, we use the Cartan scalar approach to
obtain a finite limit \cite{PaivaReboucasMacCallum1993,PaivaRomero1993}.

It is known \cite{MacCallumSkea1992} that the so called 14 algebraic
invariants (and even all the polynomial invariants of any order) are not
sufficient for locally characterizing a spacetime, in the sense that two
metrics may have the same set of invariants and be not equivalent. As an
example, all these invariants vanish for both Minkowski and plane-wave
\cite{MacCallumSkea1992} spacetimes and they are not the
same. A complete local characterization of spacetimes may be done by
the Cartan scalars. Briefly, the Cartan scalars are the components of
the Riemann tensor and its covariant derivatives (up to possibly the
$10^{\rm th}$ order) calculated in a constant frame
\cite{Paiva1995,MacCallumSkea1992,Karlhede1980,Cartan1951}. 

Therefore it is possible to establish unambiguously the local equivalence
between two given metrics by comparing their respective Cartan scalars, in
other words: Two metrics are equivalent if and only if there exist
coordinate and Lorentz transformations which transform the Cartan
scalars of one of the metrics into the Cartan scalars of the other.
It should be stressed that, although the Cartan scalars provide
a local characterization of the spacetime, global properties such as
topological defects do not probably appear in them.

In practice, the Cartan scalars are calculated using the spinorial
formalism. For the purpose here, the relevant quantities are the
Weyl spinor $\Psi_A,$ and its first covariant symmetrized derivative
$\nabla\Psi_{AB'}$, which represent the Weyl tensor and its covariant
derivative. Due to the amount of calculations, the computer algebra
systems SHEEP/CLASSI \cite{Paiva1995,MacCallumSkea1992} and MAPLE were
used throughout this section.

In order to calculate the Cartan scalars for the $\gamma$ metric, we
take the line element in spherical coordinates (eq.~(\ref{gamma-esf}))
written in the same tetrad basis used by \cite{EspositoWitten1975}. In
the $0^{th}$ order we find that the Ricci spinor and curvature scalar
vanish and the Weyl spinor satisfies the relation: $\Psi_0=\Psi_4$,
$\Psi_1=-\Psi_3$, $\Psi_2 \neq 0$. It can be easily shown that
this corresponds to a Petrov type I metric, which therefore has no
isotropies. For putting these Cartan scalars in a canonical form,
two tetrad transformations are done, which in the spinorial formalism
are given by:
\begin{equation}
\frac{1}{\sqrt{2}}\left[
\begin{array}{cc}
1 & i \\
i & 1
\end{array}
\right] ~~~ \mbox{and} ~~~
\left[
\begin{array}{cc}
A & 0 \\
0 & 1/A
\end{array}
\right]
\end{equation}
The first transformation puts the $0^{th}$ order Cartan scalars
in the form: $\Psi'_0\neq0$, $\Psi'_1=0$, $\Psi'_2\neq0$,
$\Psi'_3=0$, $\Psi'_4\neq0$. The second transformation with $A =
(\Psi'_4/\Psi'_0)^{1/8}$ gives finally: $\Psi''_0=\Psi''_4$, $\Psi''_1=0$,
$\Psi''_2\neq0$, $\Psi''_3=0$, which is the canonical form for Petrov type
I metrics.

We come out with two independent functions of the coordinates $r$ and
$\theta$ (eqs.~(\ref{CS2}) and (\ref{CS0})). 
So, up to $0^{th}$ order, the isometry group is of dimension
$4-2=2$ (where $4$ is the dimension of the spacetime). Since the metric
is independent of the coordinates $t$ and $\phi$, its isometry group is
of dimension $2$. Therefore, the $1^{st}$ order Cartan scalars will
present no new information about isometries and the Karlhede algorithm
will end in the $1^{st}$ order.

Instead of calculating the $1^{st}$ order Cartan scalars in the new basis,
for computational reasons they were calculated in the initial basis and
afterwards transformed to the new basis. Finally, to have the Cartan scalars
in the cylindrical coordinate system one has to invert the coordinate
transformation from cylindrical to spherical given by
eq.~(\ref{cil-esf})\footnote{This leads to $r = \frac{\sqrt{(z - m)^2
+ \rho^2} + \sqrt{(z + m)^2 + \rho^2} + 2m}{2} $ and $ \cos{\theta} =
\frac{\sqrt{(z + m)^2 + \rho^2} - \sqrt{(z - m)^2 + \rho^2}}{2m}$.}
and apply it to the Cartan scalars, remembering
that they transform like scalars.

In this new basis, in cylindrical coordinates, the $0^{th}$ order
Cartan scalars of the $\gamma$ metric (eqs.~(\ref{gamma-metric-ds2}) and
(\ref{gamma-metric-lambda})--(\ref{gamma-metric-R1R2})) are 
(dropping the primes):
\begin{eqnarray}
\Psi_2 &=& \frac{e^{2\lambda}}{e^{2\mu}}
\frac{m\gamma( R_1 + R_2 - 2\gamma m)}
{(R_1 + R_2 + 2m)(R_1 + R_2 - 2m)R_1 R_2}  \label{CS2} 
\\ \nonumber \\
\Psi_0 = \Psi_4 &=& - \Psi_2 
\frac{\sqrt{f^2 + g^2 }}{2 R_1 R_2 (R_1 + R_2 - 2\gamma m)}
\label{CS0}  
\end{eqnarray}
where
\begin{eqnarray}
f^2 &=& \{[(R_1 - R_2 - 2\,m)\,(R_1 - R_2 + 2\, m)\,\gamma ^{2} 
\nonumber \\
&& - (R_1 + R_2 + 2\,m)\,(R_1 + R_2 - 2\,m)](R_1 + R_2)  \nonumber \\
&& - 2(R_1 + R_2 - 6\,\gamma \,m)\,R_1\,R_2\}^{2}
\\
g^2 &=& (\gamma^2  - 1)^2\,
(R_1 - R_2)^{2}\,(R_1 + R_2 + 2\,m)\,(R_1 + R_2 - 2\,m) 
\nonumber \\
 & & (R_1 - R_2 + 2\,m)\,(R_1 - R_2 - 2\,m)
\end{eqnarray}
and $R_1$ and $R_2$ are given by eq.~(\ref{gamma-metric-R1R2}). 
The $1^{st}$ order Cartan scalars are too long and will not be shown.

Although, at first sight, the Cartan scalars seem more complicated than
the line element, a closer investigation shows that they are simpler. In
fact they depend only on the coordinates $\rho$ and $z$, while the line
element depends on these coordinates and the differentials of the four
coordinates. In other words, under coordinate transformations, the Cartan
scalars transform like scalars while the metric components transforms
like tensor components.

Due to this fact, it is easier to investigate limits using the Cartan
scalars rather than using the metric. Besides and even more important
is the fact that the metric usually has features that are due to the
non-essential coordinates (like the singularity on the Schwarzschild
horizon). On the other hand, since only the essential coordinates appear
on the Cartan scalars, they do not present such problems. So, in principle,
a coordinate system can be found which provides a well behaved limit for
the Cartan scalars while the metric still diverges.
Firstly, let us investigate the limits using $\Psi_2$, later we shall
investigate whether the other Cartan scalars share the same limits.

After a lengthy but straightforward calculation we may write:
\begin{equation}
\lim_{m\rA\infty}\Psi_2 = 2^{-(\gamma-1)}m^{2(\gamma^2-\gamma)}
\rho^{-2(\gamma^2-\gamma+1)} \gamma(1-\gamma)
\end{equation}
which is either divergent or finite depending on the value of $\gamma$. 
Nevertheless, this expression suggests that a finite limit may arise for all
values of $\gamma$ if we define a new radial coordinate $\bar\rho$ by
$m^{2(\gamma^2-\gamma)} \rho^{-2(\gamma^2-\gamma+1)} =
\bar\rho^{-2(\gamma^2-\gamma+1)}$
that is,
\begin{equation} \label{lc-coord}
\rho = 2^\beta m^\alpha \bar\rho
\end{equation}
where
\begin{equation}
\alpha=\frac{\gamma^2-\gamma}{\gamma^2-\gamma+1}
\end{equation}
and 
\begin{equation}
\beta = \frac{-\gamma}{\gamma^2-\gamma+1}
\end{equation}
The constant $\beta$ was introduced to provide the correct power of $2$
in the limiting Cartan scalar.

Indeed, noting that $-\frac{1}{3} \leq \alpha < 1$ and using
eq.~(\ref{lc-coord}) into eq.~(\ref{CS2}), a lengthy but
straightforward calculation shows that in this new coordinate system one has:
\begin{equation}
\lim_{m\rA\infty}\Psi_2 = 
\frac{1}{2}\bar\rho^{-2(\gamma^2-\gamma+1)} \gamma(1-\gamma)
\end{equation}
which is finite. Similarly, one finds that all Cartan scalars have a
finite limit in this new coordinate system. The question now is to find out to
which metric this set of Cartan scalars belongs. This is not a straightforward
task, but fortunately, calling 
\begin{equation} \label{gammasigma}
\gamma = 2\sigma
\end{equation}
and $\bar\rho =r$ we are led to following set of Cartan scalars:
\begin{eqnarray}
\psi_2 &=& (1-2\sigma)\sigma r^{-2(4\sigma^2 - 2\sigma + 1)} \\
\psi_0=\psi_4 &=& (4\sigma-1)\psi_2 \\
\nabla\psi_{01'} = \nabla\psi_{50'} &=&
\sqrt{2}(8\sigma^2 - 4\sigma + 1)(4\sigma-1)
(2\sigma - 1)\sigma r^{-3(4\sigma^2 - 2\sigma + 1)} 
\vspace{5mm} \mbox{} \\
\nabla\psi_{10'} = \nabla\psi_{41'} &=&
\sqrt{2}(4\sigma-1)
(2\sigma - 1)\sigma r^{-3(4\sigma^2 - 2\sigma + 1)} \\
\nabla\psi_{21'} = \nabla\psi_{30'} &=&
\sqrt{2}(4\sigma^2 - 2\sigma + 1)
(2\sigma - 1)\sigma r^{-3(4\sigma^2 - 2\sigma + 1)}
\end{eqnarray}
which are the Cartan scalars of the Levi-Civita
spacetime \cite{SHPS1995c}
\begin{equation} \label{Levi-Civita}
ds^2 = r^{4\sigma}dt^2 - r^{8\sigma^2-4\sigma} (dr^2 + dz^2)
- \frac{r^{2-4\sigma}}{a}d\phi^2 
\end{equation}

This shows that in this new coordinate system, the limit of the
$\gamma$ metric as $m\rA\infty$ is locally the Levi-Civita metric,
provided the radial coordinates $\bar\rho$ and $r$ are identified and the
parameter $\gamma$ divided by $2$ of the $\gamma$ metric is identified
with the density parameter $\sigma$ of the Levi-Civita metric, i.e.,
eq.~(\ref{gammasigma}) holds.

We use the word {\it locally} since the Cartan scalars provide a local
characterization of the metric. Furthermore, there is a parameter $a$
in the Levi-Civita metric which does not appear in its Cartan scalars
since it is a topological defect and can be eliminated by a coordinate
transformation. For studying the global properties of the limit one has
to investigate the metric (or the line element) directly. In fact one may ask
whether, using this new coordinate system, the Levi-Civita limit can be
obtained directly from the line element of the $\gamma$ metric.

In this new coordinate system, the $\gamma$ metric may be written as
\begin{equation} \label{gamma-metric-ds2-2}
ds^2 = e^{2\lambda}dt^2
- e^{-2\lambda} e^{2\mu} 2^{2\beta} m^{2\alpha} d\bar\rho^2
- e^{-2\lambda} e^{2\mu} dz^2
- e^{-2\lambda} 2^{2\beta} m^{2\alpha}\bar\rho^2 d\phi^2
\end{equation}
where $\lambda$ and $\mu$ are expressed in the new coordinates. The
limit of the component $g_{\bar\rho\bar\rho}$ is precisely the $g_{rr}$
of the Levi-Civita metric but the other metric components diverge. Now,
the divergences can be easily removed by similar transformations on the
coordinates $t$, $z$ and $\phi$, given by:
\begin{eqnarray}
t    &=& 2^{-\beta(\gamma^2+1)} m^{-\beta} \, \bar t \label{lc-coord-t} \\
z    &=& 2^{\beta} m^{\alpha} \, \bar z \label{lc-coord-z} \\
\phi &=& 2^{\beta\gamma^2} m^{\beta\gamma}
\, \frac{1}{\sqrt{a}} \,\bar\phi \label{lc-coord-phi} 
\end{eqnarray}
In this new coordinate system, the $\gamma$ metric becomes:
\begin{eqnarray} \label{gamma-metric-ds2-LC}
ds^2 &=& e^{2\lambda} 2^{-2(\gamma^2+1)\beta} m^{-2\beta}d\bar t^2
- e^{-2\lambda} e^{2\mu} 2^{2\beta} m^{2\alpha} d\bar\rho^2 
\nonumber \\ &&
- e^{-2\lambda} e^{2\mu} 2^{-2\beta} m^{-2\alpha} d\bar z^2
- e^{-2\lambda} 
2^{2(\beta+\gamma^2\beta)} m^{2(\alpha+\gamma\beta)}\bar\rho^2 
\frac{1}{a} d\bar\phi^2
\end{eqnarray}
and its limit is precisely the Levi-Civita metric. The only drawback of this
limit is the introduction of an infinite topological defect. In other words,
the limit of the $\gamma$ metric in this new coordinate system is the
Levi-Civita metric only locally. So we have reproduced the result we found
previously with the Cartan scalars. Whether a coordinate system for the
$\gamma$ metric exists which provides a global limit into the Levi-Civita
metric, i.e., with a finite topological defect, is still an open question.

\section{A limiting diagram for the $\gamma$ metric} 
\label{SecDiagram} \setcounter{equation}{0}
 
\begin{figure}
\setlength{\unitlength}{3ex}
\begin{center}
\begin{picture}(21,23)(0,-18)
%\graphpaper[1](0,-18)(23,21)
\put(0,0){\framebox(6,2){\shortstack{{\sc Levi-Civita} \\ Density:
$\sigma$}}}
\put(6.1,1){\vector(1,0){9.8}}
\put(10,1.3){\xs{$\sigma\rA 0$ \\ \sc(lc$^*$)}}
\put(10,-1){\xs{$\sigma\rA \frac{1}{2}$ \\ \sc(lc$^*$)}}
\put(16,0){\framebox(6,2){{\sc Minkowski}}}
\put(0,-10){\framebox(6,4){\shortstack{{\sc Gamma} \\ Length: $2m$ \\
Density: $\frac{\gamma}{2}=\sigma$ \\ Mass: $\gamma m = 2\sigma m$}}}
\put(1,-5.9){\vector(0,1){5.8}}
\put(1.2,-4){\xs{$m\rA\infty$ \\ (\sc lc$^*$)}}
\put(6.1,-5.9){\vector(3,2){9.8}}
\put(6.4,-4){\xs{$\sigma\rA 0$ \\ ({\sc lc}, $\gamma$)}}
\put(18,-5.9){\vector(0,1){5.8}}
\put(14.5,-4){\xs{$m\rA 0$ \\ ({\sc lc$^*$}, $\gamma$, {\sc s})}}
\put(20,-5.9){\vector(0,1){5.8}}
\put(20.2,-4){\xs{$m\rA\infty$ \\ ({\sc lc$^*$}, $\gamma^*$, {\sc g})}}
\put(6.1,-8){\vector(1,0){9.8}}
\put(11.5,-7.7){\xs{$\sigma\rA\frac{1}{2}$ \\ ({\sc lc}, $\gamma$)}}
\put(16,-10){\framebox(6,4){\shortstack{{\sc Schwarzschild} \\ Mass: $m$}}}
\put(1,-10.1){\vector(0,-1){5.8}}
\put(1.2,-14.5){\xs
{$\sigma\rA\infty$ \\ $m\rA 0$ \\ $2\sigma m = const.$ \\ ($\gamma$)}}
\put(0,-18){\framebox(6,2){\shortstack{{\sc Curzon} \\ Mass: $M=2\sigma
m$}}}
\qbezier[300](6,-16)(8,-7)(15.9,-0.1) \put(15.9,-0.1){\vector(1,1){0}}
\put(7.3,-13){\xs{$M\rA 0$ \\ ($\gamma$)}}
\end{picture}
\end{center}
\caption[Limiting diagram for the $\gamma$ metric.]  {Limiting diagram
for the $\gamma$ metric. In brackets are the coordinate systems
where each limit works. $\gamma$ means the original cylindrical
coordinate system for the $\gamma$ metric; {\sc lc} means the
Levi-Civita coordinate system, i.e., the $\gamma$ coordinate system
plus the coordinate transformation given by eqs.~(\ref{lc-coord})
and (\ref{lc-coord-t})--(\ref{lc-coord-phi}); {\sc s} means the usual
Schwarszchild coordinates and {\sc g} the Geroch coordinates first used
to find the Minkowskian limit of Schwarzschild.  The $^*$ means that,
the limit is local, i.e., it was taken with the Cartan scalars and/or
on the line element but a topological defect was found.}

\label{diagrama}
\end{figure}
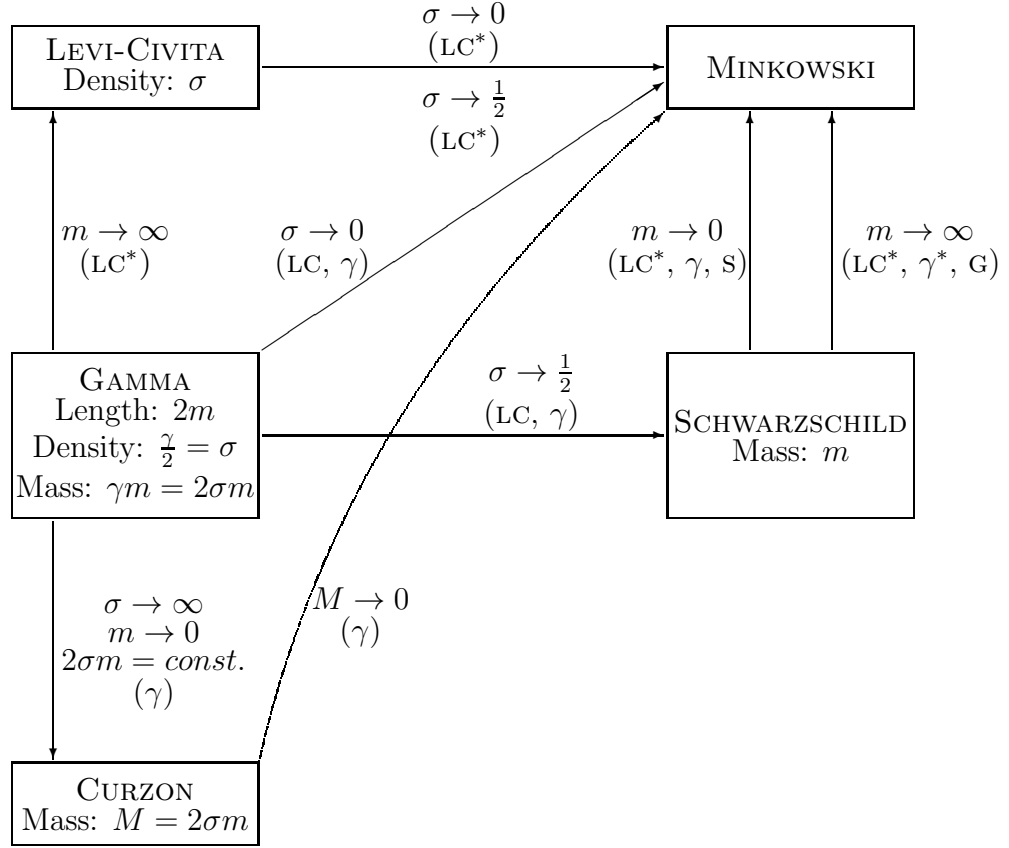

In the previous section we have shown that in the
coordinate system defined by eqs.~(\ref{lc-coord}) and
(\ref{lc-coord-t})--(\ref{lc-coord-phi}) the limit $m\rA\infty$ of
the $\gamma$ metric is locally the Levi-Civita spacetime.  We shall
now study this limit, find other limits in the coordinate systems of
the previous section and discuss some limits known in the literature in
order to build the limiting diagram for the $\gamma$ metric shown in
figure~\ref{diagrama}.

%%%%%%%%%%%%%%%%%%%%%%%%%%%%%%%%%%%%
%%%%%%%%%%%%%%%%%%%%%%%%%%%%%%%%%%%%

\subsection{Limits in the Schwarzschild coordinates}

In the usual Schwarzschild coordinates, in the
limit $m\rA 0$, the Schwarzschild line element tends to Minkowski. The
limit $m\rA\infty$ diverges. This can be easily checked by hand or from
the Cartan scalars~\cite{PaivaReboucasMacCallum1993}.

\subsection{The Geroch Limits}

In 1969 Geroch~\cite{Geroch1969} showed that in the coordinate system
(Geroch coordinates) defined by
\begin{equation}
x = r + m^{4/3}, ~~ \rho = m^{4/3} \theta,  ~~
t' = t, ~~ \varphi' = \varphi \label{s3}
\end{equation}
the limit of the Schwarzschild metric as $m\rA\infty$ is the Minkowski
spacetime.  He also presented a coordinate system where the limit is a
Kasner spacetime.  These results show that the limit of a spacetime as
some parameter goes to infinity is a coordinate dependent process.

Later, \cite{PaivaReboucasMacCallum1993} re-obtained these limits
by using the Cartan scalar technique, and extended the results
presenting new limits of the Schwarzschild metric and developing
an approach to find all limits of a given spacetime (see also
\cite{PaivaRomero1993,PaivaReboucasHallMacCallum1998,PaivaReboucasTeixeira1997})

\subsection{Limits in the $\gamma$-coordinates}

We shall call $\gamma$-coordinates the original cylindrical coordinates
used for the $\gamma$ metric (eqs.~(\ref{gamma-metric-ds2})
and (\ref{gamma-metric-lambda})--(\ref{gamma-metric-R1R2})).  Its is
known \cite{Bonnor1990} that in this coordinate system the limit
$\gamma\rA\infty$, $m\rA 0$ with $\gamma m = const.$ leads to the Curzon
metric and, as shown in section \ref{Secgamma}, the limit $\gamma\rA 1$
leads to Schwarzschild. Besides one can easily see that as $\gamma\rA
0$ the $\gamma$ metric tends to Minkowski.  The coordinate systems in
which the Curzon and Schwarzschild metrics are expressed when obtained
as limit of the $\gamma$ metric will also be called $\gamma$-coordinates.

In the $\gamma$-coordinate system, the line elements of Curzon (see
\cite{Bonnor1990}) and Schwarzschild tend to Minkowski as $m\rA 0$ (this
arises directly from the line element). Although the Schwarzschild line
element in $\gamma$-coordinates diverges as $m\rA\infty$ it can be shown
that its Cartan scalars (those of the $\gamma$ metric with $\gamma =
1$) tend to zero, which is locally Minkowski.

\subsection{Limits in the LC-coordinates}

In the previous section, starting from the $\gamma$-coordinates we defined
new coordinates by scaling $\rho$ with a coordinate transformation which
depends on $m$ and $\gamma$ (eq.~(\ref{lc-coord})). As we have then shown,
the Cartan scalars of the $\gamma$ metric in this new coordinate system
tend to the Cartan scalars of the Levi-Civita spacetime if the radial
coordinates of both metrics are identified ($\bar\rho = r$) and $\gamma =
2\sigma$. Since the Cartan scalars give a complete local characterization
of each metric, therefore, the metric corresponding to this limit is
locally the Levi-Civita metric. Rescaling also the coordinates $t$,
$z$ and $\phi$ (eqs.~(\ref{lc-coord-t})--(\ref{lc-coord-phi})) the
Levi-Civita limit was obtained directly from the line element but with
a topological defect.

This new coordinate system will therefore be called
Levi-Civita-coordinates or LC-coordinates for short, for both the
$\gamma$ metric and the Levi-Civita metric (although the metric
equivalence is only local). Coincidently, this is the
usual coordinate system for the Levi-Civita metric.

As $\gamma\rA 0$ or $\gamma\rA 1$, the $\gamma$ metric in LC-coordinates tends
to Minkowski or Schwarzschild, as can be seen directly from the line element
(\ref{gamma-metric-ds2-LC}). The last one giving Schwarzschild in
LC-coordinates.

The limits of the Levi-Civita
metric as $\sigma\rA 0$ and $\sigma\rA 1/2$ giving locally Minkowski can
be directly found from the line-element in LC-coordinates.

As $m\rA\infty$, the Schwarzschild line element in LC-coordinates diverge but
its Cartan scalars tend to $0$, i.e., locally Minkowski.

Finally, the LC-coordinate system turns out to be a new coordinate system
(Geroch coordinates is the old one) where the Schwarzschild line element
tends to Minkowski as $m\rA\infty$.  The equivalence is local since an
infinite topological defect appears. This limit can be also done with
the Cartan scalars.

\section{Conclusion} \label{SecConc} \setcounter{equation}{0}

We have seen so far that extending the length of the Newtonian
image source of the $\gamma$ metric to infinity, we arrive at the
Levi-Civita spacetime. The amazing fact is that the finite rod does
not represent the {\it real} source of the $\gamma$ metric (it is just
its Newtonian image source), whereas the infinite line singularity
is thought to be the real source of the LC spacetime. The link
between the parameters $\gamma$ and $\sigma$ ($\gamma=2\sigma$) appearing
in the limiting process is quite consistent with previous results
\cite{Geroch1969,PaivaReboucasMacCallum1993}, in the sense that the
Schwarzschild metric ($\gamma=1$) leads (locally) to Minkowski spacetime
as $m\rA\infty$ and the Levi-Civita metric ($m\rA\infty$) leads (locally)
also to Minkowski if $\sigma = 1/2$.  It should be interesting to find out
if restrictions on $\sigma$, based on the existence of timelike circular
geodesics \cite{GautreauHoffman1969} in LC ($\sigma<1/4$) do appear in
the $\gamma$ metric.

We shall now proceed to the interpretation of the limiting diagram of the
$\gamma$ metric (figure~\ref{diagrama}). In order to build this diagram, we
introduced a new coordinate system for this
metric (the LC-coordinates) and found two new limits as $m\rA\infty$:
Schwarzschild $\rA$ Minkowski in $\gamma$-coordinates and LC-coordinates
and $\gamma$ metric $\rA$ Levi-Civita in LC-coordinates.

One notices, that as it is presented, the diagram is quite consistent. 
It supports the current
interpretations of $\sigma$ as being the density in the Levi-Civita metric;
$\gamma/2$ and $2m$, respectively, as the density and the length in the
$\gamma$ metric; $m$ as the mass in the Schwarzschild solution and $M$
as the mass in the Curzon solution.

Note that from the $\gamma$ metric one can reach Minkowski either through
Levi-Civita making $m\rA\infty$ and then $\sigma\rA 1/2$ or through
Schwarzschild by making $\sigma\rA 1/2$ and then $m\rA\infty$. The limit
$\sigma\rA 0$ is similar; the difference being that, since the mass in
the $\gamma$ metric is $2\sigma m$, the limit $\sigma\rA 0$ leads to
Schwarzschild with zero mass, which is Minkowski.

This work would not be complete if we did not show the known
weakness of the diagram. The first one is that specific coordinate
systems were used. We do not know which other limits could arise if
we explored more coordinate possibilities. One away to solve this
problem would be the use of the Cartan scalar techniques developed in
\cite{PaivaReboucasMacCallum1993,PaivaRomero1993} to find all limits
of the concerning metric. The main difficulty is computational, since
in the present case too many Cartan scalars are different from zero and
they depend on more than one coordinate and parameter.

The second weakness are two other known limits of the Schwarzschild
solution found in \cite{Geroch1969} and \cite{PaivaReboucasMacCallum1993},
namely, a Kasner solution and plane wave solutions, respectively. How
do they match in this diagram is still an open problem.

Another extension of this work is finding a single coordinate
system which provides all the limits on the diagram and does not present
an infinite topological defect.  This would help the understanding of
the topological defects in the Levi-Civita metric.

\section*{Acknowledgment}

FMP gratefully acknowledges financial assistance from FAPERJ and the
Laborat\'orio de Astrof\'{\i}sica e Radioastronomia - Centro Regional Sul de
Pesquisas Espaciais (INPE/MCT - Santa Maria RS) for the kind hospitality
during the final stage of this work.

\end{document}